\documentclass[12pt,preprint]{aastex}

\shortauthors{JONES AND WEHRLE} 
\shorttitle{NGC 6251 COUNTERJET} 

\received{2002 May 13}
\begin{document}

\title{What Happened to the NGC 6251 Counterjet?} 

\author{Dayton L.~Jones} 
\affil{Jet Propulsion Laboratory, California Institute of Technology,
Mail Code 238-332,\\ 4800 Oak Grove Drive, Pasadena, CA 91109}
\email{dj@sgra.jpl.nasa.gov} 

\and

\author{Ann E.~Wehrle}
\affil{Interferometry Science Center, California Institute of Technology, 
JPL Mail Code 301-486,\\ 4800 Oak Grove Drive, Pasadena, CA 91109}
\email{aew@huey.jpl.nasa.gov} 

\begin{abstract}  
We have used the VLBA to produce a high dynamic range image
of the nucleus of NGC 6251 at 1.6 GHz and snapshot images at 5.0, 
8.4, and 15.3 GHz to search for emission from
a parsec-scale counterjet.  Previous VLBI images at 1.6 GHz  
have set a lower limit for the jet/counterjet brightness ratio near
the core at about 80:1, which is larger than expected given
the evidence that the radio axis is fairly close to the plane of the
sky.  A possible explanation is that the inner few pc of the counterjet 
is hidden by free-free absorption by ionized gas associated with
an accretion disk or torus.  This would be consistent with the 
nearly edge-on appearance of the arcsecond-scale dust disk seen
in the center of NGC 6251 by HST.  We detect counterjet emission close to 
the core at 1.6 GHz, but not at the higher frequencies.  
Given that the optical depth of free-free absorption falls off
more rapidly with increasing frequency than the optically thin 
synchrotron emission from
a typical radio jet, this result implies that the absence of 
a detectable parsec-scale counterjet at high frequencies is 
not due to free-free absorption unless the density of ionized 
gas is extremely high and we have misidentified the core at 1.6 GHz.  
The most likely alternative is a large jet/counterjet brightness ratio  
caused by relativistic beaming, which in turn requires the inner radio 
axis to be closer to our line of sight than the orientation of
the HST dust disk would suggest.  
\end{abstract}

\keywords{accretion, accretion disks --- galaxies: active --- 
galaxies: individual (NGC~6251) --- galaxies: jets --- 
galaxies: nuclei}

\section{Introduction} 

NGC 6251 (1637+826, J1632+8232, z=0.024) is an elliptical galaxy containing 
an apparently edge-on dust lane (\citet{n83}; \citet{cv97}) and a central 
stellar cusp and blue continuum light source (\citet{y79}; \citet{c93}).  
Spectroscopic observations with HST indicate the presence of a central 
black hole with a mass of $(6\pm2) \times 10^{8}\ {\rm M}_{\odot}$ \citep{ff99}.
Although NGC 6251 is located near the edge of the cluster Zw 1609.0+8212, 
recent observations by \citet{w00} show that this cluster contains at 
least three sub-clusters of galaxies that may not be physically related.
NGC 6251 is associated with one of these sub-clusters.  
The distance to NGC 6251 is $72\,h^{-1}$ Mpc, where $h = H_{\circ}\,/\,
{\rm 100}\ {\rm km\ s}^{-1}\ {\rm Mpc}^{-1}$.  The linear scale corresponding 
to 1 milliarcsec (mas) is $0.36\,h^{-1}$ pc.  The line-of-sight  
velocity dispersion of the sub-cluster is less than 300 km s$^{-1}$, 
implying that it is a poor cluster with an X-ray atmosphere temperature
of 0.7 keV \citep{w00}.  Combined with plausible densities (consistent 
with the extended X-ray luminosity measured by \citet{bw93}), this 
gas temperature is far too low to confine any part of the 
spectacularly linear (e.g., see Figure 1 in \citet{w01}) kpc-scale radio jet. 

The large-scale radio morphology of this source (\citet{w77}; \citet{w82};
\citet{p84}; \citet{j86a})  
suggests that its radio axis is close to 
the plane of the sky.  Its high declination of +83$^{\circ}$ allows 
very good (u,v) coverage by northern hemisphere VLBI arrays. 
Consequently, this is a good candidate for having a nearly edge-on
inner accretion disk which could be detected via free-free absorption 
of radiation from a parsec-scale counterjet. 

An early series of three global VLBI observations of NGC 6251 at 1.6 GHz was 
carried out during a five year period.  
The first 18-cm VLBI experiment in 1983 \citep{j86a} 
showed a one-sided jet aligned with the VLA jet and containing a knot of 
emission approximately 25 mas from the core (we assume the core 
corresponds to the strong,
unresolved peak at the eastern end of the jet).  No counterjet was detected
at a limit of 80:1 measured $\pm6$ mas from the core.  
Second and third epoch  
experiments were performed in 1985 and 1988 (\citet{j86}; \citet{jw94}) 
to look for motion of the 25-mas feature. 
No significant change in the separation between the 
core and the 25-mas knot was found (${\rm{v/c}} < 0.23\ h^{-1}$), 
although changes in jet morphology closer to the 
core were observed.
The combination of low proper motion and large jet/counterjet brightness
ratio implies, using the usual beaming model, that the 25-mas feature in the jet 
does not move with the bulk flow velocity or that the radio axis is much
closer to our line of sight than expected from the large-scale radio structure
and the nearly edge-on optical dust disk. 

An alternative explanation for the lack of a detectable counterjet near
the core is free-free absorption by ionized gas in front of the counterjet
(plausibly in the form of a parsec-scale accretion disk 
or torus oriented perpendicular
to the radio jets).  Evidence for absorption of radiation from the base of
a counterjet has been seen in several other radio sources, including 
3C84 (\citet{v94}; \citet{w94}), 
Centaurus A (\citet{j96}; \citet{tm01}), and NGC 4261 (\citet{jw97}; 
\citet{j00}; \citet{j01}).  If a similar situation exists in 
NGC 6251, the counterjet should become visible farther from the core 
at 1.6 GHz, and closer to the core at higher frequencies.  The observations
reported here were designed to detect the counterjet at 1.6 GHz, where
its intrinsic brightness is likely to be larger, and at higher frequencies
where it should be visible closer to the core.  

Our ultimate goal is to understand the extent and structure of inner  
accretion disks on parsec and sub-parsec scales.  High resolution, high
dynamic range, multi-frequency images of absorption by ionized gas 
are one of the best tools available for the study of these disks. 
NGC 6251 is one of the relatively few galaxies whose radio
axis orientation and other properties make it a good candidate for such
studies.  

\section{Observations}

We observed NGC 6251 with the VLBA and one VLA antenna for 12 hours 
on each of two days (29 and 30 May 2000).  The first day was devoted 
to 1.6 GHz to obtain the highest possible dynamic range, while on the
second day we switched between 5.0, 8.4, and 15.3 GHz with a cycle time
of 44 minutes.  This was done to provide nearly identical snapshot (u,v)
coverage at all three frequencies.    
In all cases a bandwidth of 64 MHz was recorded, with LCP polarization
at 1.6, 5.0, and 15.3 GHz and RCP at 8.4 GHz.  
The data were correlated at NRAO in Socorro, producing 16 spectral channels for
each of the 8 contiguous IF bands with an integration time of 2 seconds.
The J2000 radio position of NGC 6251, as determined by USNO astrometric 
VLBI observations, is RA = 16:32:31.97001$\pm 0.00013$,
DEC = +82:32:16.39991$\pm 0.00017$ (M.~Eubanks, private communication).  

The Los Alamos VLBA antenna was not available during either day of our
observations because of the Cerro Grande wildfire near Los Alamos.  This 
caused a loss of short and intermediate baselines that reduced our sensitivity
to very extended radio structure.  The remaining 9 VLBA antennas and one
VLA antenna produced useful data at all frequencies.  

Amplitude calibration, initial editing, and fringe-fitting were performed using
standard tasks within AIPS\footnote{The Astronomical Image Processing 
System, developed by the National Radio Astronomy Observatory.}.  After
the residual delays and rates were removed, the data were transferred 
to the Caltech program Difmap \citep{s97} for interactive editing,
self-calibration, and imaging.  
The data were averaged in time, and all 8 IFs were combined in Difmap 
during imaging.  Phase-only antenna corrections were applied during multiple
iterations of self-calibration and deconvolution until the source model
contained nearly all of the flux detected on our shortest baselines.  Then 
amplitude self-calibration was allowed, on gradually decreasing time scales.
During deconvolution, clean boxes were allowed only where emission was 
clearly visible until the final iteration.  For the final deconvolution, 
no restriction was placed on the model component locations.  

A comparison of our data with only a priori amplitude calibration applied 
and the fully amplitude self-calibrated data used to produce our final 
images showed no detectable shift in the amplitude scale caused by 
self-calibration at 1.6, 5.0, or 8.4 GHz.  At 15.3 GHz we find a small
(less than 10\%) decrease in amplitudes.  This is too small to affect 
spectral index determinations significantly.

\section{Results}

Our most complete (u,v) coverage was obtained at 1.6 GHz, and 
is shown in Figure~\ref{fig1}.   

\placefigure{fig1}   

For comparison, the least complete (u,v) coverage was obtained at 
15.3 GHz.  This is shown in Figure~\ref{fig2}.  The coverage is still
very good at this frequency thanks to the high declination of the 
source.   

\placefigure{fig2}

Figures~\ref{fig3} and \ref{fig4} show low resolution and detailed  
images of NGC 6251 at 1.6 GHz.  The low resolution image was made  
using natural data weighting and a Gaussian taper falling to 
half weight at a radial distance of 20 M$\lambda$.  All other 
images were made using uniform weighting and no taper.  
The low resolution image (Figure~\ref{fig3}) shows hints of jet emission 
more than 150 mas from the peak.  We searched for additional
emission along both the jet and counterjet directions out to
nearly 0.3 arcsec. 

\placefigure{fig3}   

\placefigure{fig4}   

Figure~\ref{fig4} shows that the jet feature approximately 100 mas 
from the peak is resolved transverse to the jet direction.  The 
morphology of this feature suggests a transverse shock, although 
its faintness precludes detailed study.  Figure~\ref{fig4} also 
shows a small change in position angle along the first 30-40 mas
of the jet.  This change in position angle, and the brightness peak 
about 25 mas along the jet, are also seen in the 1.6 GHz VLBI image of 
\citet{j86a}.  The ``25-mas peak" has nearly the same peak 
brightness relative to the core in Figure~\ref{fig4} and in 
\citet{j86a}.  This indicates that there have been no significant 
changes in this region of the inner jet since at least 1983.   
Finally, Figure~\ref{fig4} shows what apprears to be a short counterjet 
extending to about 10 mas from the peak.  
This feature was not seen in previous VLBI images at 1.6 GHz   
because it is very faint and extends on only 2-3 beamwidths from 
the brightest peak.  
The counterjet is not clearly seen above the $0.2\%$ contour level
in Figure~\ref{fig4}. 
This is below the lowest contour in the 1.6 GHz image of \citet{j86a}.  
The significance of the short counterjet will be discussed in the 
next section.  

Figure~\ref{fig5} shows our 5.0 GHz image of NGC 6251.  At the 
higher resolution of this image the ``25-mas peak" is largely 
resolved.  
The first 15 mas of the jet appears to be very straight in this 
image.   
Although there is a very faint peak about 7 mas from
the brightest peak in the direction of a counterjet, and possibly 
a very short extension of the core at low contour levels in the 
counterjet direction, neither can be considered strong evidence 
for a counterjet in Figure~\ref{fig5}.  
We have made lower resolution (tapered) images from our 5.0 GHz
data but do not detect any more extended flux in the counterjet
direction.  

\placefigure{fig5}   

Figure~\ref{fig6} shows our 8.4 GHz image.  The ``25-mas peak" is
still visible, but even more resolved.  The smooth jet breaks up
into separate peaks around 12 mas at this resolution.  There is
no evidence for a counterjet.   

\placefigure{fig6}   

Figure~\ref{fig7} shows our 15.3 GHz image.  This is our highest
resolution image, and shows the inner 5 mas of the jet plus a 
complex peak about 7 mas from the core and hints of jet emission
out to 15 mas.  There is no convincing evidence for a counterjet
at this frequency.   

\placefigure{fig7}   

Our angular resolution at 15.3 GHz is just under 0.5 mas, which corresponds
to a linear resolution of $7 \times 10^{17}$ cm (about 0.2 pc) 
for $h = 0.7$.  This is much 
greater than the gravitational radius of  
a $6 \times 10^{8}\ {\rm M}_{\odot}$ black hole, or the radius 
at which stars are tidally disrupted (which is less than the 
gravitational radius for a black hole of this mass).   
However, our linear resolution at 15.3 GHz is  
significantly smaller than the radius within which
the stellar dynamics are dominated by the black hole.  
Assuming that the black hole accounts for most of the mass distribution 
within 0.2 pc, the period of a circular orbit with this radius is  
about 400 years.  

To help compare features in the jet at different frequencies, 
Figure~\ref{fig8} shows a mosaic of model components at 5.0, 
8.4, and 15.3 GHz, all rotated clockwise by 27 degrees and 
convolved with the same restoring beam.  The models have  
been shifted horizontally to align the first peak 
along the jet, under the assumption that jet features will 
be less affected by optical depth effect than the core region. 
With this alignment, the apparent position of the core changes
with frequency in the way expected for either synchrotron 
self-absorption or free-free absorption.  In either case we 
can see emission from closer to the true base of the jet 
at higher frequencies.  

\placefigure{fig8}

The observing parameters (frequency, resolution, contour levels,
etc.) for each of our images are summarized in Table~\ref{tab1}.

\placetable{tab1}  

We also made images at adjacent pairs of frequencies convolved 
with the same restoring beam to look at spectral index variations 
within the source.  For each pair of frequencies, the common 
restoring beam was a circular gaussian whose width was equal 
to or only slightly less than the true resolution at the lower
frequency.  Comparing the matched-resolution images at 1.6 and 
5.0 GHz (4 mas restoring beam) showed that the brightness peak has an 
inverted spectrum while the jet along position angle $296^{\circ}$ 
has a steep spectrum, especially around 10 and 25 mas from the 
peak.  The spectral index $\alpha$ (defined as S$_{\nu} \ \propto \ 
\nu^{\alpha}$) varies from about +0.5 at the peak to -0.6 near 
10 and 25 mas down the jet.  Comparing matched resolution images
at 5.0 and 8.4 GHz (1.3 mas beam) gave a nearly flat spectum 
for the brightness peak, a steep spectrum for the jet around 
5-7 mas from the peak, and complex variations in spectral index
farther along the jet.  Finally, our matched resolution images
at 8.4 and 15.3 GHz (0.8 mas beam) again showed a flat spectrum
for the peak and a very steep spectrum for the jet about 2-3
mas from the peak.  Farther out (to about 10 mas) the jet 
spectrum is less steep, with $\alpha \approx -0.6$.  The flat 
or inverted spectrum of the core could be caused by either 
free-free absorption or synchrotron self-absorption.

\section{Discussion} 

\citet{s01} see a very short ($\approx 1$ mas) counterjet in their 
VLBI images of NGC 6251 at 5 and 15 GHz. 
Our 5.0 GHz image also shows a possible short extension in the
counterjet direction from the brightest peak, but we do not
consider this a reliable feature because of its faintness and
the fact that its extent is no greater than one beamwidth. 
Note that our angular resolution at 5 GHz is not as good as 
that of \citet{s01}, who had data from the HALCA spacecraft of
the VSOP mission in addition to their ground-based VLBI array.  
The counterjet in \citet{s01}'s 5 GHz image appears at an 
unexpected position angle (not parallel to the main jet) 
and only at the higher resolution provided by the HALCA data. 
We do not see any flux along the position angle of the \citet{s01} 
counterjet in our lower resolution but higher dynamic range 
image at 5.0 GHz.  

We see no evidence for a counterjet in our 15.3 GHz image.  
At this frequency our angular resolution is similar to that 
obtained by \citet{s01}, who see a short extension from 
the peak in the counterjet direction.  Our image at 15.3 GHz 
also appears to show a 
very short and faint extension in the counterjet direction, 
but we do not consider this feature to be reliable because 
of its faint and barely resolved nature.  

The most surprising result from our observations is the lack of
a clearly detected parsec-scale counterjet at 5.0, 8.4, and 15.3 GHz. 
A large optical depth due to free-free absorption at 15.3 GHz would 
imply a high electron density or a long line-of-sight path length  
through the ionized material ($n^{2}\,l,> 10^{27}\ {\rm cm}^{-5}$, 
where $n$ is electron density per cm$^{-3}$ and $l$ is the path
length in cm).  Neither situation would be
consistent with the apparent short counterjet seen at 1.6 GHz.  
Thus, we conclude that free-free absorption is not the primary
reason for the lack of a detectable counterjet at high frequencies. 
The other obvious possibility is that the counterjet is hidden by 
the effects of Doppler boosting.  This, in turn, implies that the
radio axis is close enough to our line of sight that jet/counterjet
brightness ratios (R) greater than 100 are possible.  For a continuous
jet, the angle between the jet axis and our line of sight must 
be $\theta \le 47^{\circ}$ for R $\ge 100$.  

NGC 6251 has the third largest linear size (nearly 2 Mpc for $h = 0.7$)
out of the 84 large 
angular size radio galaxies studied by \citet{l01}.  It is the 
largest FR I galaxy \citep{fr74} in this sample.
This suggests that it is unlikely to be seen highly foreshortened.  
\citet{p84} detected a radio counterjet extending $\sim 130$ arcsec
from the core in low resolution VLA images.  The jet/counterjet 
brightness ratio at $\pm 100$ arcsec from the core is R $\sim 40$, 
but at $\pm 240$ arcsec R $\ge 200$.  
At 8.4 GHz we measure R $> 128$ at $\pm 2.5$ mas.  
Thus, the brightness ratio varies signficantly (and not monotonically) 
with distance from the core.  The Doppler boosting explanation then 
requires large-scale variations in bulk velocity or radio axis 
orientation (unless the jet is clumpy and intrinsically asymmetric).   
Another difficulty with Doppler boosting is that  
\citet{g96} estimate an equipartition Doppler factor \citep{r94}  
of 1.4 for the core of NGC 6251, and an inverse Compton 
Doppler factor of only 1.0.  
If either of these estimates applies to
the bulk motion of material in the VLBI jet, it is not possible to 
obtain brightness ratios as large as 128 for any orientation.  
This difficulty can be removed if the core Doppler factors are
treated as lower limits (see \citet{g97}).

Thus, the simplest versions of free-free absorption and Doppler
boosting are insufficient to explain our multi-frequency images
of NGC 6251.  We now consider more complex versions of these
two basic mechanisms.  Free-free absorption could be responsible
for the lack of detectable counterjet emission if the extension
we see at 1.6 GHz is not actually from a counterjet but is 
actually highly absorbed core emission.  In this case, we would 
identify the brightest peak in our 1.6 GHz image as the region
where the jet appears from behind a geometrically thick disk 
or torus of ionized gas, which hides the counterjet and partially
hides the core.  At higher frequencies the free-free absorption 
is insufficient to hide the core emission, but can still hide
the counterjet due to the longer line-of-sight path length. 
Two ways to test this possibility would be 1) VLBI imaging at 
2.3 GHz, which would fill in the largest fractional gap in 
our current frequency coverage, and 2) phase-referenced VLBI 
observations to allow unambiguous registration of images at 
1.6 GHz and higher frequencies.  

The accretion disk expected to surround the central black hole
in a radio galaxy like NGC 6251, with an intermediate 
radio luminosity and no very strong broad emission lines, should 
be geometrically thick according to the model of \citet{m01}.   
A thick disk would be necessary to hide the core at low 
frequencies, given the likely orientation of the radio axis
at an intermediate angle to our line of sight.  
This model also predicts that very active phases, characterized 
by a large increase in jet power, may occur for short periods 
every few thousand years.  Interestingly, such behaviour could
explain the quasi-periodic chains of bright knots seen along 
the kpc-scale radio jet in NGC 6251 \citep{s00}.  

The major problem with this model is the high density of ionized
gas required to hide the counterjet at high frequencies.  The 
HI absorption detected in front of the radio nucleus by the VLA, 
visual extinction by the dust disk imaged with HST, and the X-ray  
absorption of the core determined from ASCA and ROSAT spectra  
all imply column densities of $~\sim 10^{21}\ {\rm cm}^{-2}$ 
\citep{w01}.  That is, the HST disk alone 
can provide all of the absorption needed to explain the radio,
optical, and X-ray measurements.  A large additional absorbing 
column density on parsec scales would be difficult to reconcile 
with the observed degree of core X-ray absorption unless most 
of the X-ray emission comes from a region much larger than a parsec.  

Our remaining alternative is a more complicated version of the
Doppler boosting model discussed earlier.  To account for the
variations in jet/counterjet brightness ratio as a function of
distance from the core, the jet direction and/or velocity must
change significantly.  Both are possible; is there evidence 
for either?
The dust disk and the inner ionized gas disk studied by \citet{ff99} 
are not coplanar, and neither is perpendicular to the radio jet axis.
The dust disk is more nearly orthogonal to the radio axis in 
position angle, but the disk models developed by \citet{ff99} 
imply that the normal to the dust disk is much farther from our
line of sight ($\sim 76^{\circ}$) than is the normal to the 
inner gas disk ($\sim 35^{\circ}$).  Therefore, if the radio 
axis is closer than $\sim 50^{\circ}$ to our line of sight it
will be more nearly perpendicular to the plane of the inner gas
disk than to the plane of the dust disk.  
Figures \ref{fig9} and \ref{fig10} show the orientations of 
the radio jet, gas disk, and dust disk as a function of 
distance from the core.  

\placefigure{fig9}

\placefigure{fig10}

The non-coplanar nature of the nuclear disks on different 
scales suggests that the angular momentum of accreting gas
changes over long time scales, and consequently the direction
of the central black hole's angular momentum vector (and radio
jet) may change with time.  Evidence for this on time scales
of $10^{6}$ years or longer is seen in the large-scale S-shape 
symmetry of the radio source \citep{w77}.  However, the inner two arcminutes
of the jet are extremely straight, so any short-term variations
in jet direction must be mainly in a plane perpendicular to the plane
of the sky.  This is possible, but seems contrived.  

Variations in jet velocity are perhaps more plausible.  
Unfortunately it is difficult to get more than a lower limit
for the bulk velocity from multi-epoch VLBI observations.  
\citet{a01} use numerical simulations to show that regions of
enhanced radio emission in a relativistic jet may have differing
apparent motions, as is frequently seen in VLBI observations, 
even when the bulk velocity along the jet is nearly constant.  
In particular, they show that bright regions
near the base of a jet may move along the jet more slowly than
regions farther down the jet, with regions close to the core 
remaining nearly stationary.  Such behavior is seen is some 
VLBI monitoring observations of superluminal sources (e.g., \citet{g01}), 
although the opposite behavior is also seen in other sources.  
What is clear is that existing observations have difficulty 
in determining unambiguous jet flow velocities, or changes in 
velocities.  Consequently we cannot rule out this explanation 
for the variations in jet/counterjet brightness in NGC 6251. 

The observed VLBI properties of NGC 6251 can be compared with  
those of other intermediate or low luminosity radio galaxies 
for which high quality VLBI images exist.  In particular, we 
consider 3C84, NGC 4261, M87, and Centaurus A, all of which are 
FR-I radio sources that have been observed with VLBI at 
multiple frequencies and epochs.  There is strong
evidence for free-free absorption of radio emission from 
parsec-scale counterjets in 3C84 \citep{w00a}, NGC 4261 \citep{j01}, 
and Centaurus A \citep{j96}.  However, 
the situation in M87 is less clear because a parsec-scale 
counterjet has not been detected in this source at any 
frequency \citep{b99}.  It 
is plausible that the radio axis in M87 is aligned closer to 
our line of sight than is the case for NGC 4261 or Centaurus A, 
and as a result we might expect the jet/counterjet ratio to be
larger for M87 than for the other sources.  However, 3C84 also 
has clearly detected counterjet emission and is believed to be
aligned close to our line of sight.  In this case the counterjet
emission is detectable only because of the unusually large 
dynamic range achieved in VLBI images of 3C84.  

Thus, we can speculate that sources whose radio axes are 
oriented far from our line of sight are able to show evidence 
of free-free absorption near their cores, but sources 
aligned closer to our line of sight are likely to have 
undetectable counterjets due to relativistic beaming.   
Consequently the effect of free-free absorption cannot be 
seen in the more aligned sources.  NGC 4261 and Centaurus A 
clearly fall into the first category, while M87 and NGC 6251 
appear to fall into the second (more aligned) category.

\section{Conclusions} 

Free-free absorption does not appear to be responsible for the 
lack of a detectable parsec-scale counterjet at high frequencies 
in this source, 
unless high electron densities or path lengths are 
invoked and the short counterjet we see at 1.6 GHz is actually 
the (highly absorbed) core. 
The most likely alternative is a large jet/counterjet brightness
ratio caused by relativistic beaming, which in turn requires the
inner radio axis to be closer to our line of sight than the 
orientation of the HST dust disk would suggest.  In this case
the smaller jet/counterjet brightness ratio seen farther from
the core would be evidence for a gradual increase in the jet 
angle to our line of sight or a decrease in the bulk velocities
in the jets.

\acknowledgements
The Very Long Baseline Array and the Very Large Array are facilities
of the National Radio Astronomy Observatory, which is operated by
Associated Universities under a cooperative agreement with the 
National Science Foundation.  This research has made use of NASA's 
Astrophysics Data System Abstract Service.  
AEW~acknowledges support from the NASA Long Term 
Space Astrophysics Program.
We thank the anonymous referee for several helpful suggestions. 
This research was carried out at the Jet 
Propulsion Laboratory, California Institute of Technology, under 
contract with the National Aeronautics and Space Administration.

\newpage

\figcaption{(u,v) coverage obtained at 1.6 GHz.  Note that some
annular gaps in coverage appear because of the loss of the Los 
Alamos VLBA antenna.  However, the coverage remains very good.  
\label{fig1}}

\figcaption{(u,v) coverage obtained at 15.3 GHz.  Even though 
observations on this day were switched between three frequencies,
the resulting coverage is still quite good. \label{fig2}}

\figcaption{VLBA image of NGC 6251 at 1.6 GHz, showing emission  
the jet out to nearly 200 mas.  The coutours are -0.05, 
0.05, 0.1, 0.2, 0.4, 0.8, 1.6, 3.2, 6.4, 12.8, 25.6, 51.2, and 99\% 
of the peak brightness, which is 294 mJy/beam.  The restoring beam is 
8.00 $\times$ 7.76 mas, with the major axis along position
angle $-23.2^{\circ}$. \label{fig3}}

\figcaption{VLBA image of NGC 6251 at 1.6 GHz, showing the inner
few tens of pc in more detail.  The contours are -0.1, -0.06, 0.06, 
0.1, 0.2, 0.4, 0.8, 1.6, 3.2, 6.4, 12.8, 25.6, 51.2, and 99\% of the peak
brightness, which is 227 mJy/beam.  The restoring beam is 
4.14 $\times$ 3.92 mas, with the major axis along position 
angle $-26^{\circ}$.  Note the short extension from the peak
in the counterjet direction.  \label{fig4}}

\figcaption{VLBA image of NGC 6251 at 5.0 GHz.  The coutours 
are -0.1, 0.1, 0.2, 0.4, 0.8, 1.6, 3.2, 6.4, 12.8, 25.6, 52.1, 
and 99\% of the peak brightness, which is 246 mJy/beam.  The 
restoring beam is $1.35 \times 1.30$ mas, with the major axis 
along position angle $-7.2^{\circ}$. \label{fig5}}

\figcaption{VLBA image of NGC 6251 at 8.4 GHz.  The contours 
are -0.2, -0.1, 0.1, 0.2, 0.4, 0.8, 1.6, 3.2, 6.4, 12.8, 25.6, 
52.1, and 99\% of the peak brightness, which is 277 mJy/beam. 
The restoring beam is $0.87 \times 0.81$ mas, with the major
axis along position angle $27.3^{\circ}$. \label{fig6}}

\figcaption{VLBA image of NGC 6251 at 15.3 GHz.  The contours
are -0.2, 0.2, 0.4, 0.8, 1.6, 3.2, 6.4, 12.8, 25.6, 51.2, and 99\% of the 
peak brightness, which is 271 mJy/beam.  The resotoring beam 
is $0.49 \times 0.46$ mas, with the major axis along position 
angle $8.7^{\circ}$. \label{fig7}}

\figcaption{Mosaic of VLBI model components at 5.0 GHz (top),
8.4 GHz (middle), and 15.3 GHz (bottom).  All three models have
been rotated clockwise by 27 degrees and convolved 
with the same restoring beam, a circular Gaussian
with a diameter of 0.75 mas.  Note that this beam is only 
slightly more than half the proper beam size at 5.0 GHz.  It is
close to the proper beam at 8.4 GHz, and about 50\% larger than
the proper beam at 15.0 GHz.  The contours are 0.25, 0.5, 1, 2, 
4, 8, 16, 32, 64, and 99\% of the peak brightness. 
The tick marks along the axes are separated by 1.3 mas. \label{fig8}} 

\figcaption{Measured position angles of the radio jet, the 
stellar rotation axis, the minor axes of the gas and dust disks,
and the isophotal minor axis of NGC 6251.  The vertical sizes 
of the labeled regions indicate the published position angle  
uncertainties, and the horizontal sizes indicate the approximate
range of angular scales over which the position angles were 
determined.  \label{fig9}} 

\figcaption{Derived inclination angles for the radio jet, gas 
disk, and dust disk in NGC 6251.  As in Fig.~\ref{fig9}, the 
vertical extent of each region indicates the published 
uncertainty and the horizontal extent indicates the range 
of angular scales for which the (model dependent) inclination 
angle was determined.  \label{fig10}} 
\newpage
\hoffset=-3.3cm
\begin{tabular}{|c|c|c|c|c|l|} 
\hline
\multicolumn{6}{|c|}{Parameters of VLBA Images of NGC 6251} 
   \\ \hline\hline
\multicolumn{1}{|c|}{Figure}
& \multicolumn{1}{|c|}{Freq.}
& \multicolumn{1}{|c|}{Beam} 
& \multicolumn{1}{|c|}{P.A.} 
& \multicolumn{1}{|c|}{Peak}
& \multicolumn{1}{|c|}{Contours} 
\\ 
\multicolumn{1}{|c|}{No.}
& \multicolumn{1}{|c|}{(GHz)}    
& \multicolumn{1}{|c|}{(mas)} 
& \multicolumn{1}{|c|}{(deg.)}
& \multicolumn{1}{|c|}{(mJy/beam)} 
& \multicolumn{1}{|c|}{(\% of peak)} 
\\ \hline
3 & 1.6 & $8.00 \times 7.76$ & -23 & 294 & -0.05, 0.05, 0.1, 0.2, 0.4, 0.8, 
1.6, 3.2, 6.4, 12.8, 25.6, 51.2, 99.0 \\ \hline 
4 & 1.6 & $4.14 \times 3.92$ & -26 & 227 & -0.1, -0.06, 0.06, 0.1, 0.2, 
0.4, 0.8, 1.6, 3.2, 6.4, 12.8, 25.6, 51.2, 99.0 \\ \hline
5 & 5.0 & $1.35 \times 1.30$ & -7 & 246 & -0.1, 0.1, 0.2, 0.4, 0.8, 
1.6, 3.2, 6.4, 12.8, 25.6, 51.2, 99.0 \\ \hline
6 & 8.4 & $0.87 \times 0.81$ & 27 & 277 & -0.2, -0.1, 0.1, 0.2, 0.4, 0.8, 
1.6, 3.2, 6.4, 12.8, 25.6, 51.2, 99.0 \\ \hline
7 & 15.3 & $0.49 \times 0.46$ & 9 & 271 & -0.2, 0.2, 0.4, 0.8, 1.6, 
3.2, 6.4, 12.8, 25.6, 51.2, 99.0 \\ \hline 
8 & 5,8,15 & $0.75 \times 0.75$ & 0 & 326 & 0.25, 0.5, 1.0, 2.0, 4.0, 
8.0, 16.0, 32.0, 64.0, 99.0 \\ \hline 
\hline
\label{tab1} 
\end{tabular}
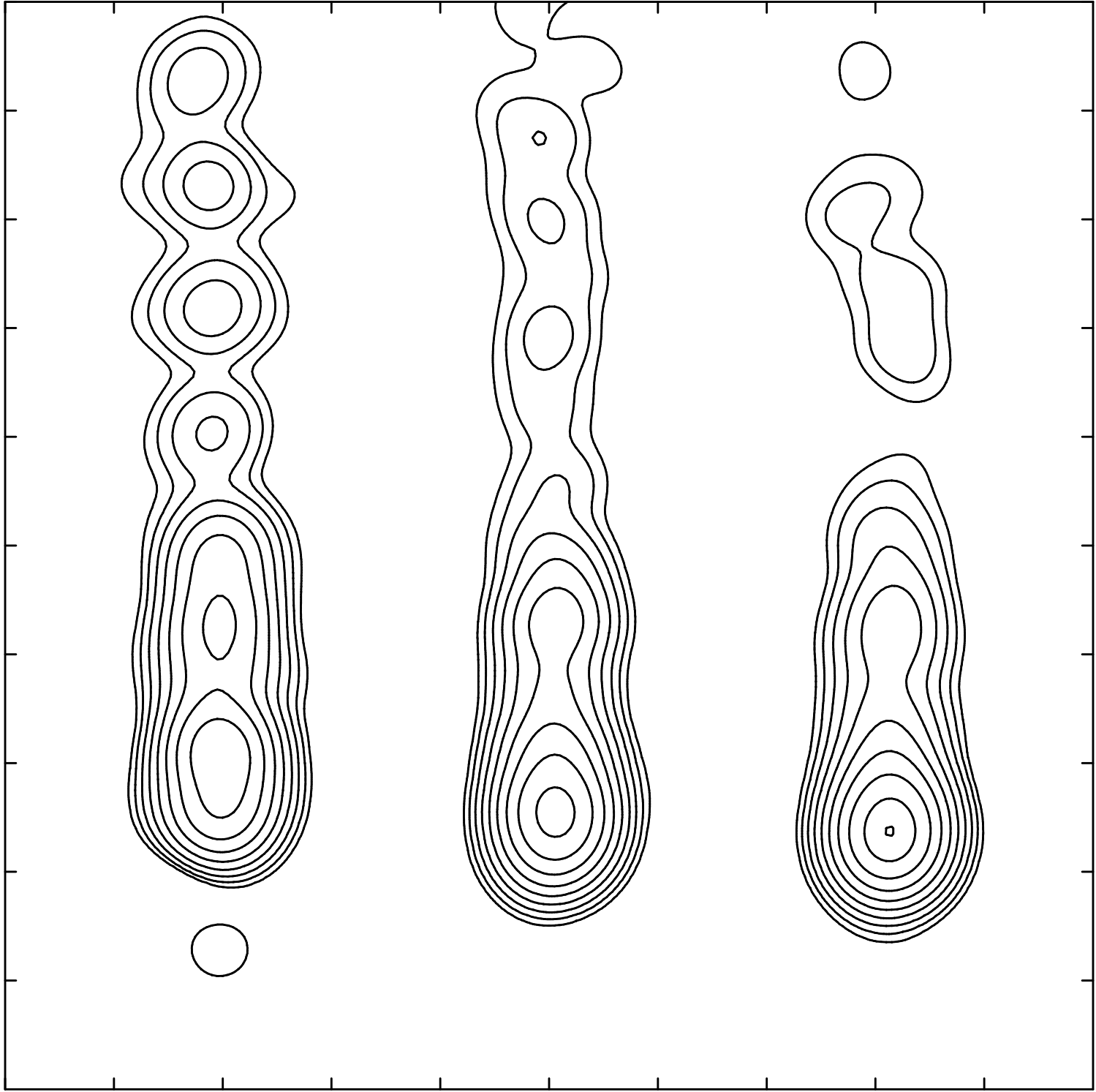
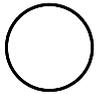
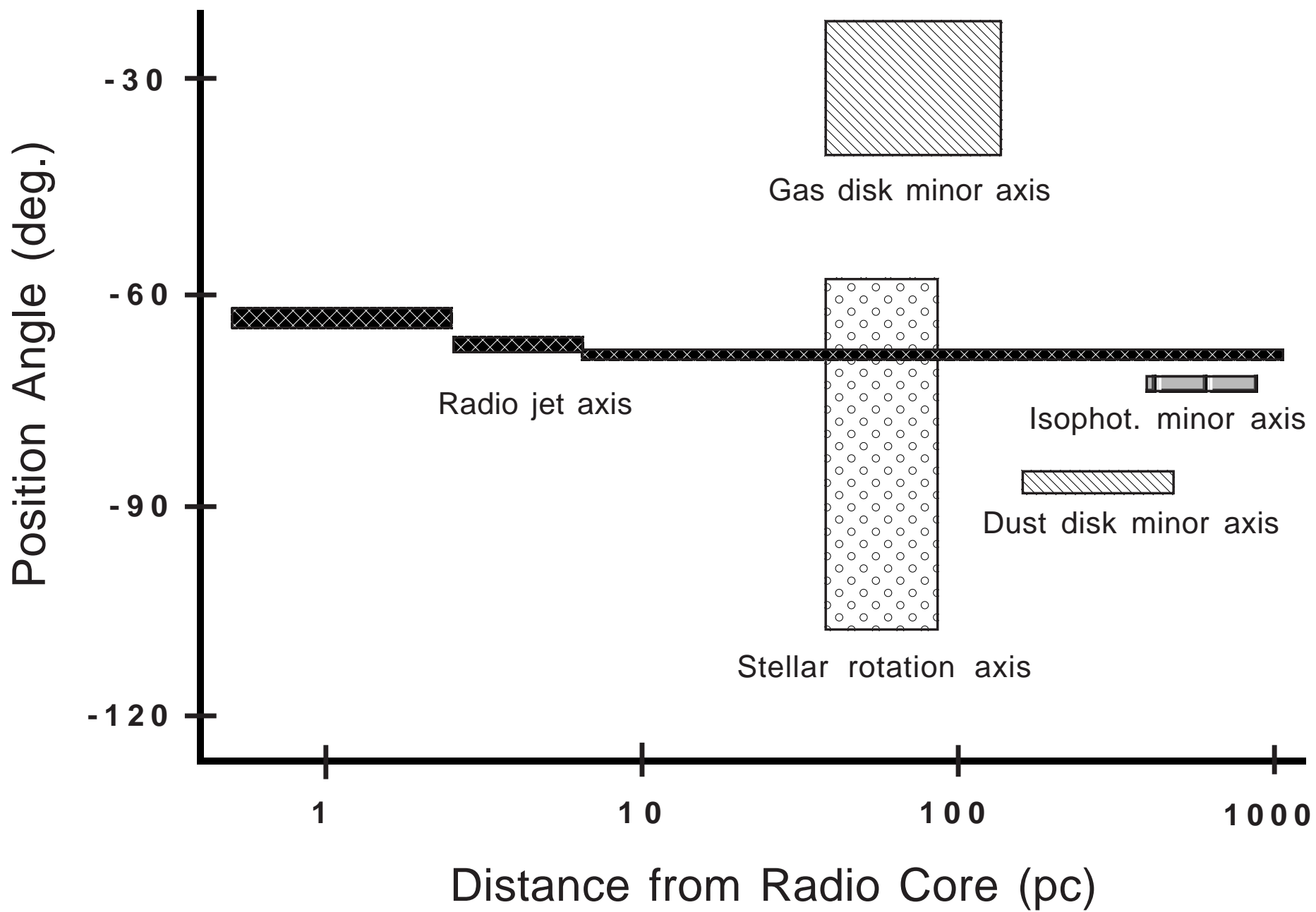
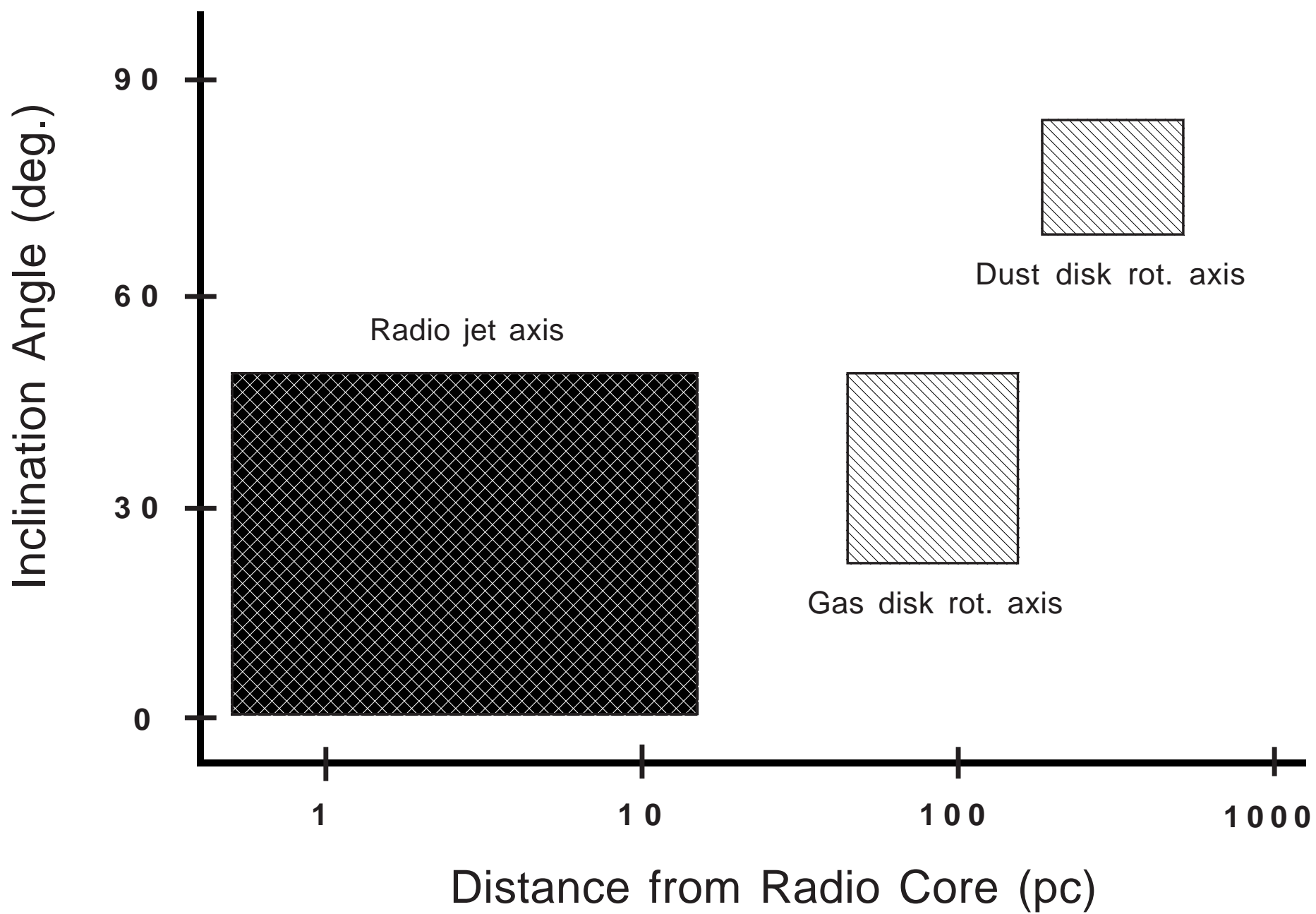
\end{document}